\documentclass[11pt,a4paper,twoside]{article}

	\usepackage{amssymb}
	\usepackage{amsmath}
	\usepackage{bm}
	\usepackage{graphicx}
	\usepackage{subfigure}
	\usepackage{hyperref}
	\usepackage{color}
	\usepackage{revsymb}
	\usepackage[font=small,labelfont=bf]{caption}
	\usepackage[title]{appendix}
	\usepackage[cbgreek]{textgreek}

	\usepackage{booktabs}
	\usepackage{array}
	\usepackage{enumitem}
	\usepackage{cite}
	\usepackage[affil-it,auth-lg]{authblk}
	\usepackage{hyperref}
	\usepackage{eso-pic}

\newcommand{\LA}{\left\langle}
\newcommand{\RA}{\right\rangle}
	
\newcolumntype{K}[1]{>{\centering\arraybackslash}p{#1}}

\begin{document}

\title{Thermodynamic origin of quantum time-energy uncertainty relation}

\author[1]{Zacharias Roupas\footnote{\href{mailto:Zacharias.Roupas@bue.edu.eg}{Zacharias.Roupas@bue.edu.eg}}}
\affil[1]{Centre for Theoretical Physics, The British University in Egypt, Sherouk City 11837, Cairo, Egypt} 

\date{\vspace{-5ex}}


\maketitle

\begin{abstract}
	The problem of time is a notable obstacle towards the recognition of quantum theory as the ultimate fundamental description of nature. Quantum theory may not be complete if founded upon classical notions. Louis de Broglie, seeming to be more or less convinced about the ontology of his proposed matter waves, tried to develop a theory of sub-quantum degrees of freedom relying on statistical thermodynamics. He realized a quantum particle as a fluctuating dense corpuscle formed via non-linear effects from a sub-quantum medium. A wave on the medium guides the vibrating corpuscle. He argued that an intrinsic clock of a quantum particle is related to its Brownian motion at the sub-quantum level. This led him to conjecture a relation between the de Broglie clock frequency $m c^2/h$ and its implicit temperature, which equals that of the surrounding sub-quantum medium. About the same time, Mandelbrot was the first to derive in a classical setting a thermodynamic uncertainty relation between energy and temperature, that was, coincidentally or not, anticipated by Bohr and Heisenberg in the first years of development of quantum theory. We show here that, when the de Broglie temperature-time conjecture is assumed, the thermodynamic temperature-energy uncertainty relation leads to the quantum time-energy uncertainty relation. 
\end{abstract}

\section{Introduction}

The theorem of Pauli \cite{1980gpqm.book.....P}\footnote{Page 63.}, which forbids the definition of a Hermitian operator conjugate to the energy, highlights the problem of time in quantum mechanics. Thus, Pauli concludes, time must necessarily be an ordinary number, not a dynamical operator. Several attempts have been made to overcome Pauli's theorem and define a time operator \cite{auletta2000foundations,2017EJPh...38c5402B,Khorasani:2016uoe,Yau:2020ooa} without a definite, fully general result.
Still, the quantum time-energy uncertainty relation may receive at least eight interpretations, enumerated for example in \cite{2002tqm..conf...69B}, because of this ambiguity of integrating a classical notion within a quantum framework. 	
Also related is the problem of irreversibility. The equations of Quantum Mechanics are reversible, yet the measurement process and decoherence are irreversible \cite{auletta2000foundations}, reminding the problem of irreversibility in Statistical Mechanics.
The status of Quantum Mechanics as the ultimate fundamental theory is not secured, while, eventually, the notion of time may be proven to be its Achilles' heel. 

Recalling the process of deposition of classical mechanics from the status of the fundamental theory in the early twentieth century, one may recognize the monumental role of Statistical Mechanics and Thermodynamics towards this development, emphasized by the works of Planck and Einstein. This may not have come as a surprise, since the advantage of Statistical Mechanics is that it allows drawing general quantitative conclusions about systems with different underlying microscopic details. Therefore, it is the ideal arena for inspecting the collective behavior of unknown degrees of freedom. Louis de Broglie, considering seriously the possibility of a sub-quantum reality since the beginnings of the development of quantum physics at 1927 \cite{1927_deBroglie}, called forth the powerful framework of statistical physics on his side \cite{1964_article_deBroglie,deBroglie_book_1964,1970FoPh....1....5D,1987_deBrogie}. In 1964, he presented a theory of the ``hidden thermodynamics of particles'' \cite{1964_article_deBroglie,deBroglie_book_1964} in which a quantum particle is realized as a dense corpuscle of energy, vibrating as a clock and being subject to Brownian fluctuations driven by a sub-quantum medium through the exchange of heat. He conjectured an equality between the particle's internal temperature and clock's frequency.

About the same time de Broglie revived this theory \cite{1964_article_deBroglie} (that he initially abandoned following the Solvay Conference of 1927), Mandelbrot, as in 1956 \cite{Mandelbrot_1956}, proved the thermodynamic uncertainty relation between energy and temperature in a purely classical framework. He argued that energy and temperature, as well as canonical and microcanonical ensembles as physical conditions, are complementary. Such a result was anticipated since much earlier by 
Bohr and Heisenberg \cite{Bohr_1932}, who conjectured complementarity of energy and temperature in the initial stages of the development of quantum physics in order to clarify and argue in favor of quantum complementarity. 
In Bohr's words according to Heisenberg \cite{Bohr_foundations_1985iii}
``\textit{As soon as I know the temperature, then the concept of energy has no meaning}.''
Since Mandelbrot's derivation, the energy-temperature uncertainty relation has been confirmed in many different contexts \cite{Rosenfeld_1961,Lindhard_1986lqt..conf...99L,1988JPCS...49..679S,Phillies_10.1119/1.13583,Lavenda_1987IJTP...26.1069L,Feshbach_1987PhT....40k...9F,Uffink1999,Velazquez_2009MPLB...23.3551V,Velazquez-Curilef_NHC_2009,Velazquez_2009JPhA...42i5006V,Falcioni_10.1119/1.3563046,Velazquez_2012AnPhy.327.1682V,Miller+Anders_2018NatCo...9.2203M}, while the notion of thermodynamic complementarity is nowadays better understood \cite{1989PhT....42a..71M,Falcioni_10.1119/1.3563046,2015AnPhy.363...48M,2017PhR...709....1P}.

A deep connection between the Heisenberg's uncertainty principle and thermodynamics was revealed recently by the intriguing observation that violation of the quantum uncertainty principle implies a violation of the second law of thermodynamics \cite{2013NatCo...4E1670H}. 

Here, we shall apply the de Broglie temperature-time conjecture. Starting from the thermodynamic energy-temperature uncertainty relation, we shall derive the quantum time-energy uncertainty relation, suggesting its statistical origin.

In the next section we shall review the de Broglie theory of a hidden thermodynamics and discuss the temperature-time conjecture. In Section \ref{sec:therm_unc} we shall review the thermodynamic uncertainty relation. In Section \ref{sec:derivation} we shall derive the quantum time-energy uncertainty relation from the thermodynamic one and we shall conclude our results in the last section.

\section{The de Broglie clock and temperature-time conjecture}\label{sec:temp_conj}
		
		The de Broglie's theory of matter waves underlies modern quantum mechanics and relativistic quantum field theory. As de Broglie has described in several occasions \cite{1964_article_deBroglie,deBroglie_book_1964,1970FoPh....1....5D,1987_deBrogie} his relation, that revealed the wave nature of matter and is fundamental for quantum mechanics,
		\begin{equation}\label{eq:lambda_deBroglie}
				p = \frac{h}{\lambda}, 
		\end{equation}
	was in fact inspired by special relativity. He remarked that the frequency of a clock transforms under Lorentz boosts $\text{\textbeta}$ (beware not to confuse $\text{\textbeta}$ with $\beta$ that shall denote inverse temperature) as
	\begin{equation}\label{eq:nu_clock}
		\nu_{\rm clock} = \nu_{{\rm clock},0}\sqrt{1 - \text{\textbeta}^2}
	\end{equation} 
inversely than the frequency of a plane monochromatic wave, which transforms as
	\begin{equation}
		\nu_{\rm wave} = \frac{\nu_{{\rm wave},0}}{\sqrt{1-\text{\textbeta}^2}},
	\end{equation} 
where the subscript zero denotes the proper reference frame. 
The former equation (\ref{eq:nu_clock}) is simply the retardation of clocks in motion. The latter equation may be explained as follows. Consider a standing wave $\psi \propto e^{i2\pi \nu_0 t_0}$ perceived as such in the reference frame of an observer $O_0$. Assume another observer $O$ which moves with velocity $V = c\,\text{\textbeta}$ with respect to $O_0$ and recall the Lorentz transformation for time $t_0 = (t + Vx/c^2)/\sqrt{1-\text{\textbeta}^2}$. The standing wave in the frame $O_0$ is perceived as a wave in the $O$ frame traveling in the opposite direction with phase velocity $V_{\rm ph} = -c/\text{\textbeta}$  \cite{1964_article_deBroglie}
\begin{equation}
	\nu_0 t_0 = \nu_0 \frac{t + Vx/c^2}{\sqrt{1-\text{\textbeta}^2}} = 
	\frac{\nu_0}{\sqrt{1-\text{\textbeta}^2}}\left(t - \frac{x}{V_{\rm ph}}\right)
	=\nu \left(t - \frac{x}{V_{\rm ph}}\right),
\end{equation}
and with frequency $\nu = \nu_0/\sqrt{1-\text{\textbeta}^2}$.

Louis de Broglie further remarked that if one attributes to the corpuscle, representing the matter part of a quantum particle, an internal clock with frequency
\begin{equation}
	\nu_{{\rm clock},0} = \frac{m_0 c^2 }{h},
\end{equation}
suggested by the fundamental quantum relation
\begin{equation}
		E = h\nu,
\end{equation}
and if one assumes that in the proper system of the corpuscle, the wave that is associated with it is a stationary wave of frequency $\nu_{{\rm clock},0}$, then all of the formulas of wave mechanics, and most notably relation (\ref{eq:lambda_deBroglie}), are directly deduced (see \cite{deBroglie_book_1964}). While moving in the wave, the corpuscle has an internal vibration which is constantly in phase with that of the wave as follows from the guidance formula (e.g. see \cite{1987_deBrogie})
\begin{equation}\label{eq:pilot_vel}
V_{\rm particle} = - c^2\frac{\nabla \varphi}{\partial_t\varphi} \overset{\rm c\ll 1}{\longrightarrow} - \frac{\nabla \varphi }{m_0},
\end{equation}
where $\varphi$ is the phase of the wave and the first equation is relativistic while the limit corresponds to the non-relativistic case.
		
		Nowadays, atom interferometry \cite{Lan554} and electron channeling experiments \cite{2017EJPh...38c5402B,2019CaJPh..97...37B} suggest that the de Broglie clock is truly an intrinsic property of massive particles.
		In addition, Roger Penrose suggested that the passage of time is observable only as soon as the universe contains massive particles \cite{penrose2011cycles}\footnote{Section 2.3.}.

However, what is rarely to our knowledge mentioned is that de Broglie continued his thought extending the analogy between quantities with similar relativistic transformations to thermodynamics. He noticed that the difference between the relativistic transformation formulas for energy  
$
	E = E_0/\sqrt{1-\text{\textbeta}^2}
$
and for heat or temperature
$
	Q = Q_0\sqrt{1-\text{\textbeta}^2}
$,
$
	T = T_0\sqrt{1-\text{\textbeta}^2}
$
is totally analogous to the difference, that had impressed him formerly, between the formula of transformation of the frequency of a wave and that of the frequency of a clock. This led him hypothesize the existence of an intrinsic temperature of the corpuscle coupled with the clock
\begin{equation}\label{eq:deBroglie}
	kT = h\nu_c,
\end{equation}
where we denote $\nu_c = \nu_{\rm clock}$. 
In de Broglie's view, a quantum particle may be modeled by a corpuscle whose motion is not only guided by a pilot wave, via equation (\ref{eq:pilot_vel}), on a surrounding medium, that de Broglie compares to the Bohm-Vigier sub-quantum medium \cite{PhysRev.96.208}, but the corpuscle is ``\textit{incorporated into the wave as if it were part of its structure}'' \cite{1970FoPh....1....5D}. It is subject to Brownian motion in a medium with temperature given by (\ref{eq:deBroglie}).

Working in the framework of relativistic wave mechanics, de Broglie, realizes that the trajectory of the particle is bended by a quantum force that is equal to the gradient of the quantity
\begin{equation}
	M_0  = \sqrt{m_0^2 + \frac{\hbar^2}{c^2}\frac{\Box |\psi|^2}{|\psi|^2}},
\end{equation}
that he calls the `variable proper mass' (we denote $\Box = (1/c^2)\partial_t^2 - \nabla^2$). He finds \cite{1964_article_deBroglie}\footnote{Page 11.} $\LA M_0\RA = m_0$. Therefore, the constant proper mass $m_0$ that is usually attributed to the particle appears to us as the mean value of the true instantaneous proper mass $M_0$, which fluctuates. Since it is an internal property of the particle, de Broglie attributes a quantity of heat $\delta Q_0$ to any variation $\delta M_0 c^2$. Recognizing the Lagrange function $\mathcal{L} = -M_0 c^2 \sqrt{1-\text{\textbeta} ^2}$ he is able to write (e.g. \cite{1964_article_deBroglie})
\begin{equation}
	\delta Q = - \delta \mathcal{L},
\end{equation}
where the variations are supposed to be with respect to $M_0$. He manages to arrive to a formal correspondence between the maximum entropy principle and the principle of least action \cite{1987_deBrogie}
\begin{equation}\label{eq:S_deBroglie}
	\frac{S}{k} = \frac{I}{h}, \quad 
	S = S_0  - \frac{M_0}{m_0}
\end{equation}
where $I = \int \mathcal{L} dt$, the entropy $S_0$ is supposed to contain only the entropy of the particle not depending on $M_0$, and variations are with respect to $M_0$. 
According to de Broglie \cite{1970FoPh....1....5D} ``\textit{The principle of least action is but a particular case of the second law of thermodynamics.
The privileged role, whose paradoxical character has been underlined by
Schr\"odinger, that present quantum mechanics attributes to plane monochromatic waves and to stationary states of quantified systems can be explained by the fact that they correspond to entropy maxima, not because the other states are nonexistent, but only because they are of a lesser probability.}'' 

Later, Silva and Pereira recovered the de Broglie's theory as a result of the wavy character of a quantum particle and hence of the many corresponding degrees of freedom \cite{Silva+Pereira_1970}, that are themselves responsible for the entropy and temperature of the particle. In a similar development, very recently Gr\"ossing \cite{2010Entrp..12.1975G}\footnote{See equation (3.3.17).} has proved equation (\ref{eq:deBroglie}) in the context of sub-quantum thermodynamics as the balance equation of an oscillator (particle) immersed in the heat bath of the environment (sub-quantum medium).

Let us comment further that a similar relation between temperature and time is also hinted by quantum field theory and quantum gravity. 
The quantization of a field $\Phi$ may be incorporated with the path-integral formulation of the transition amplitude 
$
	\LA \Phi_2,t_2|\Phi_1,t_1\RA = \int_1^2 \mathcal{D}\Phi e^{ \frac{i}{\hbar}I[\Phi]},
$
where the path-integral runs over all field configurations with $\Phi_1 = \Phi(t_1)$, $\Phi_2 = \Phi(t_2)$ and $I[\Phi]$ is the functional of the action. It also holds
$
	\LA \Phi_2,t_2|\Phi_1,t_1\RA = \LA\Phi_2 \right.| e^{- \frac{i}{\hbar} H \tau} |\left.\Phi_1\RA,
$
where $H$ is the Hamiltonian operator and $\tau = t_2 - t_1$. Feynman \& Hibbs \cite{feynman1965quantum}\footnote{See page 273.} realized that if $\tau$ is the period of the field such that $\Phi_1 = \Phi_2$, then a canonical ensemble at temperature $T$ may naturally be attributed to the field. The substitution 
\begin{equation}\label{eq:wick}
	\frac{i}{\hbar}\tau \rightarrow \beta,
\end{equation}
with $\beta = 1/kT$ (not to be confused with the velocity $\text{\textbeta}$), gives
\begin{equation}
	\text{Tr} e^{-\beta H} = \int_1^2 \mathcal{D}\Phi e^{ \frac{i}{\hbar}I[\Phi]},
\end{equation}
Gibbons \& Hawking \cite{PhysRevD.15.2752} used this property towards their development of Euclidean Quantum Gravity and derived the Bekenstein-Hawking law for black hole entropy. 
In these formulations as well as in modern quantum field theory \cite{Peskin:1995ev}, a correspondence between quantum systems and statistical mechanics is established by a Wick rotation $t \rightarrow i t$ and the substitution (\ref{eq:wick}). 

Here, we conjecture de Broglie's relation (\ref{eq:deBroglie}), remaining agnostic on the origin of the intrinsic temperature and the possible theory underlying quantum mechanics, recognizing that the rest of de Broglie's theory is not essential for our result. 
However, we remark that in our perspective any quantum particle does implicitly hide information about its ontological state and therefore should be characterized by internal entropy and therefore temperature.
Then, one might recognize a correspondence between temperature and time realizing that any passage of time requires the succession of events. Such a succession is impossible for zero temperature (e.g. photons according to (\ref{eq:deBroglie})) while it accelerates with increasing temperature.

\section{Thermodynamic Uncertainty Relation}\label{sec:therm_unc}

Since Mandelbrot's first derivation of a thermodynamic uncertainty relation between energy and temperature \cite{Mandelbrot_1956}, many authors have re-derived it \cite{Rosenfeld_1961,Lindhard_1986lqt..conf...99L,Phillies_10.1119/1.13583,Lavenda_1987IJTP...26.1069L,Feshbach_1987PhT....40k...9F,1988JPCS...49..679S,Uffink1999,Velazquez_2009MPLB...23.3551V,Velazquez-Curilef_NHC_2009,Velazquez_2009JPhA...42i5006V,Falcioni_10.1119/1.3563046,Velazquez_2012AnPhy.327.1682V,Miller+Anders_2018NatCo...9.2203M}
in different contexts ranging from classical fluctuation theory to statistical inference.
The common ground of these derivations is the perspective of statistical thermodynamics, as incorporated by Landau \cite{1980_landau_statistical}\footnote{Chapter XII.}, where instead of microscopic states, one works directly with probability distributions over macroscopic variables. This treatment was originally introduced by Einstein \cite{1905AnP...322..549E,einstein2011investigations}.

\subsection{Classical fluctuation theory}\label{sec:fluc_theory}
Schl\"ogl \cite{1988JPCS...49..679S} managed to derive a generally valid energy-temperature uncertainty relation using Einstein's postulate in the context of statistical thermodynamics and fluctuation theory. Let us briefly review this derivation here, while we also refer the interested reader to recent developments \cite{Falcioni_10.1119/1.3563046,2015AnPhy.363...48M,2017PhR...709....1P}. 

	 Consider a small system and its environment at fixed temperature $T_o = \beta_0^{-1}$ and generalized forces $F_{0,i}$ (e.g. pressure). Assume each of these (small system and environment) is in internal thermal equilibrium but not in equilibrium with each other (this may occur at timescales shorter than the relaxation timescale of the combined system). Denote $S_{\rm tot}$ the total entropy of the total system (both the small system and its environment), capital $Y$ a state of the total system and small $y$ a state of the small system, while the subscript zero denotes equilibrium states. Einstein's postulate inverts Boltzmann's formula $S \propto k\ln W$ assigning probabilities to $Y$ in terms of their entropy. It may be written as 
\begin{equation}
	W(Y) = W(Y_0) e^{\frac{1}{k}(S_{\rm tot}(Y) - S_{\rm tot}(Y_0))}
\end{equation}
where $W(Y)$ are interpreted as the relative frequencies that the states $Y$ occur, i.e. they are related to the probability of fluctuations. 
The entropy fluctuation $S_{\rm tot}(Y) - S_{\rm tot}(Y_0)$ can be calculated in terms of the minimal amount of the work needed to restore the equilibrium state $Y_0$. This can be expressed with respect to the state variables $y$ of the small system
\begin{equation}
	S(Y(y)) - S(Y_0(y)) = -\beta_0 \left(E(y) + \sum_i F_{0,i} X_i (y) \right) + S(y)
\end{equation}
where $S(y)$ is the entropy of the small system and $X_i$ the generalized displacements (e.g. volume). Then, the quantity
\begin{equation}
	P = \frac{W(Y)}{W(Y_0)}.
\end{equation}
expresses the probability of a fluctuation to occur.
Considering $E,X_i$ as the independent state variables we get
\begin{equation}\label{eq:prob} 
	P(E,\{X_i\}) =  e^{-\frac{1}{k}(\beta_0(E + \sum_i F_{0,i} X_i ) + \frac{1}{k}S(E,\{X_i\})}.
\end{equation}
The quantity 
\begin{equation}
	\beta(E,\{X_i\}) \equiv \frac{\partial \ln P(E,\{X_i\})}{\partial E} + \beta_0
\end{equation}
may be identified as the generalized out-of-equilibrium inverse temperature, since it satisfies
\begin{equation}
	\beta(E,\{X_i\}) = \left( \frac{\partial S}{\partial E}\right)_{\{X_i\}}
\end{equation}
The general inequality 
\begin{equation}
	\Delta E \Delta \beta \geq |{\rm Cov(E, \beta)}| \equiv 
	\LA \left(E - \LA E\RA \right) \left(\beta - \LA \beta\RA \right)\RA,
\end{equation}
assuming mean values over the probability (\ref{eq:prob}) gives the thermodynamic uncertainty relation
\begin{equation}\label{eq:Schlogl}
	\Delta E\Delta \beta \geq k .
\end{equation}

At this point, let us remark that a few authors \cite{Feshbach_1987PhT....40k...9F,1988PhT....41e..93K,Uffink1999} had questioned in the past the notion of complementarity between energy and temperature on the basis that as they argued temperature fluctuations cannot be defined or are zero in the microcanonical ensemble. 
The recent developments \cite{Falcioni_10.1119/1.3563046,2015AnPhy.363...48M,2017PhR...709....1P} that clarified the notion of temperature fluctuations for isolated systems, we believe, have settled the issue. Intuitively, one may recognize that even in an isolated system the kinetic energy of the constituents fluctuates, as we shall also discuss in the next subsection. 
  
In particular, the derivation above can be generalized for isolated systems assuming that the small system and the environment constitute an isolated system, divided into many subsystems. 
The subsystems are in internal equilibrium at a timescale $t_{\rm sub}$ much shorter than the relaxation timescale of the total system $t_{\rm tot}$, while they fluctuate at an intermediate timescale $t_{\rm fl}$. Such a partition to subsystems may always be achieved because the relaxation timescale depends on the size of the system, while recent developments suggest that the fluctuation timescale is bounded by the energy transfer \cite{2015PhRvL.114o8101B,PhysRevX.10.021056,2020NatPh..16...15H,2020NatPh..16.1211N}. This timescale hierarchy $t_{\rm sub} \ll t_{\rm fl} \ll t_{\rm tot}$ allows for considering temperature fluctuations of subsystems, even in an isolated system at equilibrium. One has just to realize that a thermal equilibrium state is defined in all cases only at certain sufficiently large timescales, while its definition fails at shorter ones, when fluctuations of its subsystems (that are physically realized in grand canonical or canonical ensembles) emerge.
Such a theory of equilibrium fluctuations has been already rigorously developed by Mishin \cite{2015AnPhy.363...48M} and we shall not discuss it further here.

\subsection{Statistical inference}\label{sec:stat_inf}

The first derivation of the thermodynamic uncertainty relation between energy and temperature was performed by Mandelbrot \cite{Mandelbrot_1956,1989PhT....42a..71M}. We refer the interested reader to \cite{Falcioni_10.1119/1.3563046} for a clarifying review.
Mandelbrot's starting point is again statistical thermodynamics, defining a probability over macroscopic states. However, he proceeds then by incorporating methods of estimation theory, as follows. 

In Mandelbrot's analysis, the energy $E$ of a canonical thermodynamical system is considered a random variable whose probability takes the form
\begin{equation}
	P(E|\beta) = \sigma (E) e^{-\frac{1}{k}E\beta} Z^{-1}(\beta),
\end{equation}
where $\sigma$ is the density of states and $Z$ the partition function. Mandelbrot considers the statistical estimator $\hat{\beta}$ of $\beta$, that has a well defined variance bounded by the Cram\'er-Rao inequality (e.g. see \cite{bickel_doksum}), which for an unbiased estimator $\LA \hat{\beta}\RA = \beta$ reads
\begin{equation}\label{eq:Cramer-Rao}
	(\Delta \hat{\beta})^2 = \LA (\hat{\beta} - \beta)^2\RA \geq I_F^{-1}.
\end{equation}
The quantity $I_F$ is the Fisher's information, that equals in our case
\begin{align}
	I_F(\beta) &= \int \frac{1}{P(E|\beta)} \left(\frac{\partial P(E|\beta)}{\partial \beta}\right)^2 dE  \nonumber \\
	&= \int \frac{1}{k^2}\left( E + \frac{\partial k\ln Z}{\partial \beta} \right)^2 P(E|\beta) dE = \frac{1}{k^2} \left(\LA E^2\RA - \LA E\RA^2 \right)= \frac{1}{k^2}(\Delta E)^2.
\end{align}
Substituting into the Cram\'er-Rao inequality we get directly the thermodynamic uncertainty relation
\begin{equation}\label{eq:mandelbrot}
		\Delta E \Delta \hat{\beta} \geq k.
\end{equation}
In this uncertainty relation the $\Delta \hat{\beta}$ expresses the uncertainty in estimating $\beta$.
Mandelbrot replied \cite{1989PhT....42a..71M} to the critic against energy-temperature complementarity \cite{Feshbach_1987PhT....40k...9F,1988PhT....41e..93K} (on the basis that in the microcanonical ensemble there can be no temperature fluctuations), discussed in the previous section, by arguing that in an isolated system it is actually the temperature that is imperfectly defined and not its fluctuations. He viewed the uncertainty in temperature as a measure of the disagreement between Gibbs and Boltzmann definitions of temperature in finite systems. He considered these to be definitions of estimators and not of the respective physical quantity. In Ref. \cite{1989PhT....42a..71M} he concluded ``\textit{An imperfectly defined microcanonical temperature with a well-defined fluctuation may at first seem strange, but there should be no insurmountable difficulty in achieving consensus on its behalf.}'' Mandelbrot's approach seem to agree with our concluding paragraph of previous section. In the microcanonical ensemble, we argued that temperature may approximately be defined only at sufficiently large timescales, when the kinetic energy fluctuations (induced by the two-body potential energies), can be ignored. 

Especially enlightening is the approach of Falcioni etal. \cite{Falcioni_10.1119/1.3563046}. These authors conclude that because the mean value of the kinetic energy per particle $K$ is proportional to the
temperature, it might be inferred directly that the fluctuations of $K$ are related to the fluctuations of the temperature. Therefore, they argue, the random variable $K$ may be used as a statistical estimator of the temperature $\hat{T}$ with ${<(\Delta K)^2>} \propto {<(\Delta \hat{T})^2>} $.

In our perspective, the Mandelbrot uncertainty relation (\ref{eq:mandelbrot}), although conceptually differrent, reflects the fluctuation induced uncertainty relation (\ref{eq:Schlogl}), because the uncertainty of temperature measurements reflects the temperature fluctuations.

\section{Thermodynamic Derivation of Time-Energy Uncertainty Relation}\label{sec:derivation}

We consider here the implications of de Broglie conjecture (\ref{eq:deBroglie}) with respect to the quantum uncertainty principle.
The thermodynamic uncertainty relation
\begin{equation}\label{eq:thermal_u}
	\Delta E\Delta \beta \geq k,
\end{equation}
combined with de Broglie conjecture
\begin{equation}
	\frac{\beta}{k} = \frac{t_c}{h}
\end{equation}
directly implies a lower bound to the uncertainty of the de Broglie clock period $t_c = 1/\nu_c$
\begin{equation}\label{eq:dt_clock}
	\Delta t_c \geq \frac{h}{\Delta E}.
\end{equation}

Measurements of time intervals shorter than the period of the clock do not make physical sense. Such measurements are definitely bound by (\ref{eq:dt_clock}). 
Let us therefore consider measurements of time intervals $t>t_c$.
We shall define the variation of $t$ by use of a phase 
\begin{equation}\label{eq:dphi}
	\phi \equiv 2\pi \int^t \nu_c(\tilde{t}) d\tilde{t} \Leftrightarrow
	\frac{d\phi (t)}{dt} = 2\pi\nu_c (t).
\end{equation}
It follows that
\begin{equation}
	k T = \hbar \dot{\phi}.
\end{equation}
Since $\phi$ does not depend explicitly on $t$, we define the uncertainty in time by the change of $\phi$ as (see also \cite{messiah1965quantum}\footnote{Section VIII.13.})
\begin{equation}
	\Delta t = \frac{\Delta \phi}{\left| d\LA \phi\RA /dt\right|},\quad t > t_c.
\end{equation}
According to this Mandelstam-Tamm \cite{Mandelstam+Tamm_1946} type of definition, the time uncertainty expresses the time required for $\phi$'s statistical distribution to be appreciably modified.

Now, we have that provided $\Delta\nu < \nu_c$ it holds by a Taylor expansion
\begin{equation}
	\LA \frac{1}{\nu_c}\RA = \frac{1}{\LA\nu_c\RA} + \mathcal{O}(\Delta\nu_c ^2/\LA\nu_c\RA^3).
\end{equation}
It follows that
\begin{equation}\label{eq:delta_nu_c}
	\Delta \left(\frac{1}{\nu_c(t)} \right)  = \frac{\Delta \nu_c}{\LA\nu_c\RA^2} + \mathcal{O}(\Delta\nu_c ^2/\LA\nu_c\RA^3).
\end{equation}
It also holds
\begin{equation}
	\Delta \int^t \nu_c(\tilde{t}) d\tilde{t}  \geq \int^t \Delta \nu_c(\tilde{t}) d\tilde{t}
	\geq t_c \Delta\nu_c(t),\quad \text{for } t \geq t_c,
\end{equation}
since by definition $\Delta\nu_c \geq 0$ and $\LA \Delta\nu_c \RA = 0$.
Substituting (\ref{eq:delta_nu_c}) into the latter inequality we get
\begin{equation}\label{eq:nu_c-ineq}
	\Delta \left(\frac{1}{\nu_c(t)} \right) \leq \frac{\Delta \int^t \nu_c(\tilde{t}) d\tilde{t} }{\LA \nu_c\RA}.
\end{equation}
This is equivalent to
\begin{equation}\label{eq:t_c-ineq}
	\Delta t \geq \Delta t_c.
\end{equation}
Inequality (\ref{eq:t_c-ineq}) is physically reasonable expressing that the precision by which we measure time cannot be higher than the precision of the clock which defines the passage of time, namely $\nu_c$.

Substituting (\ref{eq:t_c-ineq}) into (\ref{eq:dt_clock}), which follows from the thermodynamic uncertainty (\ref{eq:thermal_u}), we get Heisenberg's uncertainty relation for time and energy
\begin{equation}\label{eq:Heisenberg}
	\Delta E\, \Delta t \geq h.
\end{equation}

\section{Conclusion}\label{sec:discussion}

Despite its amazing, unprecedented success in a wide range of fields from high energy to atomic physics and chemistry, quantum theory presents some weaknesses as a physical theory. These do not only refer to the many possible interpretations of the theory (e.g. \cite{jammer1974}) and the problem of reconciliation with gravity, but also to the difficulty of integrating the notions of irreversibility and time into the theory in a self-consistent and efficient manner. Therefore, it seems justified to investigate for possibilities that do not consider Quantum Mechanics as an ultimate fundamental theory, but look for underlying degrees of freedom, focusing especially on the notion of time. This is a road that Louis de Broglie, among others, followed \cite{deBroglie_book_1964}.

It is a prediction of the special theory of relativity that time ticks only for massive particles, while massless ones appear frozen in time.
Louis de Broglie suggested a theory, in which a quantum massive particle possesses an implicit clock, whose ticking frequency depends on its temperature. Such a temperature arises because the particle is realized as a dense turbulent corpuscle of energy, formed via non-linear effects from surrounding sub-quantum degrees of freedom, with which it exchanges heat. The particle is subject to Brownian motion at this temperature.

However, we have remarked here that there exists an uncertainty relation between the energy and temperature of a system, derived originally by Mandelbrot \cite{Mandelbrot_1956}. It reflects a type of complementarity between energy and temperature. Any attempt to define the energy via a microcanonical ensemble, renders temperature indefinite. Likewise, putting the system in a heat bath renders its energy indefinite. 
Using only the de Broglie conjecture between temperature and time (\ref{eq:deBroglie}), and not necessarily any other aspect of his theory, we have shown that the thermodynamic energy-temperature uncertainty relation gives rise to a time-energy uncertainty relation. 

There is an important issue we did not address that may lead us or others to future developments. If the time-energy uncertainty relation is of statistical nature as suggested here it stands on different grounds than the quantum versions. Meaning that designing experiments which could verify or falsify the thermodynamic origin of time-energy uncertainty may be possible. This requires further investigation.
Another possible future development is the reformulation and update of the forgotten de Broglie's theory under the light of modern developments not only in quantum mechanics and information theory, but also in the fields of quantum gravity and holography.

\bibliography{Roupas_Thermal-Uncertainty_arxiv_2021}
\bibliographystyle{myunsrt}

\end{document}